\documentclass[12pt]{article}

\usepackage{graphicx,amsmath,bm}

\begin{document}

\title{{\Large {\bf Low-Lying States of $^{6}$He and $^{6}$Be in
a Nodal Surface Structure Analysis}}}
\author{{Yu-xin Liu$^{1,2,4,}$\thanks{Corresponding author}, Jing-sheng Li$^1$,
and Cheng-guang Bao$^{3,4}$} \\
\normalsize{$^1$ Department of Physics, Peking University, Beijing
100871, China}\\
\normalsize{$^2$ Institute of Theoretical Physics, Academia Sinica,
Beijing 100080, China} \\
\normalsize{$^3$ Department of Physics, Zhongshan University,
Guangzhou
510275, China} \\
\normalsize{$^4$ Center of Theoretical Nuclear Physics, National
Laboratory of }\\
\normalsize{Heavy Ion Accelerator, Lanzhou 730000, China} }
\maketitle

\begin{abstract}
The low-lying states of the light nuclei $^6$He and $^6$Be are
studied. Using the inherent nodal surface(INS) analysis approach,
we deduce the quantum numbers and the spatial symmetries of the
low-lying states with positive parity and negative parity of the
two nuclei. The energy spectrum obtained agrees well with the
experimental data.
\end{abstract}

\bigskip

\bigskip



\newpage

It is known that, when the number of nucleons $A$ in a nucleus is
5-10, the convergence of shell model calculations is usually poor,
while in the cluster model it is complicated to include many
different cluster configurations. Although many attempts have been
made to investigate the nuclei with $A \approx 5$-10 , no general
approach has been established because of the complexity due to the
many degrees of freedom, for instance, six-nucleon systems have 15
spatial degrees of freedom. Among these nuclei, $^6$He has
received considerable attention since studying such a nucleus at
the neutron drip line can further refine our understanding of the
nucleon-nucleon interaction. $^{6}$He, consisting of two protons
and four neutrons, has a level structure which has been
investigated over many years in a number of theoretical
calculations. However, the existing literature concerns mainly the
ground states and a few
resonances\cite{Six1,Six2,Six3,Six4,DRE97,Three}. Even now, little
is known about the spectroscopy and energy level scheme of $^6$He
and $^6$Be. To shed more light on the situation, we make use of a
new approach, namely the inherent nodal surface structure analysis
approach, to study the effects of the inherent symmetries. It has
been shown that, by investigating the nodal structure of the
few-body wave functions, one can obtain certain important features
of the wave functions and the energy spectra before actually
solving the Schr\"odinger or Faddeev equation\cite{He4,SWBL}. The
properties of the 6-nucleon systems have been studied in this new
approach, and the isospin $T=0$ energy level scheme of $^6$Li has
been deduced\cite{SWBL}, as an example. In this paper we will
extend this approach to extract the qualitative characteristics of
the low-lying states of $^6$He and $^6$Be with isospin $T=1$.

It has been shown in Ref.\cite{He4} that the low-lying states of
$^4$He are dominated by the component of total orbital angular
momentum $L=0$, while the resonances below the 2n+2p threshold are
ruled with $L=1$. The quite large excitation energies of the
resonances ($\geq 20$~MeV) indicate that the increase of $L$ may
lead to a great increase in energy. Furthermore, the internal
wave-functions (the ones relative to a body-frame) of all the
states below the 2n+2p threshold do not contain nodal surfaces.
This fact implies that the excitation of internal oscillation
takes a very large energy. Since 6-nucleon systems have comparable
size and weight with the 4-nucleon ones, it would be reasonable to
assume that the $L=0$ nodeless component will dominate the
low-lying spectra of 6-nucleon systems. The success in describing
the positive parity energy spectrum of $^{6}$Li\cite{SWBL}
indicates that such an assumption is quite practical.

Let $\Psi $ be an eigenstate of a quantum system, ${\cal{A}}$
denote a geometric configuration, in some cases ${\cal{A}}$ may be
invariant to a specific operation $\hat{O}$, we have then
\begin{equation}
 \hat{O}\Psi ({\cal{A}})=\Psi (\hat{O}{\cal{A}})=\Psi
 ({\cal{A}})\, .
\end{equation}
For example, when ${\cal{A}}$ is a regular octahedron (OCTA, see
Fig.~1) for a 6-body system, ${\cal{A}}$ is invariant to a
rotation about a 4-fold axis of the OCTA by $90^{\circ }$ together
with a cyclic permutation of the particles 1, 2, 3 and 4.
According to the representations of the operation on $\Psi $
(e.g., rotation, space inversion, and permutation), Eq.~(1) can
always be written in a matrix form and appears as a set of
homogeneous linear equations. It is apparent that, whether there
exists nonzero solution of $\Psi({\cal{A}})$, in other word,
whether the state $\Psi$ is accessible to the configuration
${\cal{A}}$, depends on the inherent symmetric property of the
configuration. The symmetry imposes then a very strong constraint
on the eigenstate so that the $\Psi $ may be zero at ${\cal{A}}$.
It indicates that there may exist a specific kind nodal surface.
Since such kind nodal surface is imposed by the intrinsic symmetry
of the system (fixed at body-frames) and independent of the
dynamical property at all, one usually refers it as inherent nodal
surface (INS)\cite{He4,SWBL,Bao}.

The INS appears always at geometric configurations with certain
geometric symmetry. For a 6-body system, the OCTA is the
configuration with the strongest geometric symmetry. Let us assume
that the six particles form an OCTA,  $k^{\prime}$ is a 4-fold
axis of the OCTA, the particles 1, 2, 3 and 4 form a square
surrounding $k^{\prime}$, $R_\delta ^{k^{\prime }}$ denotes a
rotation about $k^{\prime}$ by an angle $\delta $ (in degree),
$P(ijk)(P(ijkl))$ denotes a cyclic permutation of the particles i,
j, k (i, j, k, l), the OCTA\ is evidently invariant to operations
\begin{equation}
\hat{O}_1=P(1432)R_{-90}^{k^{\prime }} \, ,
\end{equation}
\begin{equation}
\hat{O}_2 = P(253)P(146)R_{-120}^{oo^{\prime }}\, .
\end{equation}
Let $P_{ij}$ denote an interchange of the locations of particles
$i$ and $j$, $\hat{I}$ stand for a space inversion, along the same
way as discussed above, we know that the OCTA is also invariant to
the operations
\begin{equation}
\hat{O}_3 = P_{14}P_{23}P_{56}R_{180}^{i^{\prime }}\, ,
\end{equation}

\begin{equation}
\hat{O}_4 = P_{13}P_{24}P_{56}\hat{I}
\end{equation}

On the other hand, we can generally express an eigenstate of a
6-nucleon system with given total angular momentum $J$, parity
$\pi$, and isospin $T$ as
\begin{equation}
\Psi =\sum_{L,S,\lambda}\Psi _{LS\lambda}
\end{equation}
where $S$ is the total spin, and the $\Psi_{LS \lambda}$ can be
given as
\begin{equation}
\Psi _{LS\lambda}=\sum_{i M M_{S} }C^{JM_J}_{LM,SM_S}
F_{LSM}^{\lambda i}\chi _{SM_S}^{\widetilde{\lambda}i} \, ,
\end{equation}
where $M$ is the $Z$-component of $L$, $F_{LSM}^{\lambda i}$ is a
function of the spatial coordinates, which is the $i^{th}$ basis
function of the $\lambda -$representation of the permutation group
S$_6$, $\chi _S^{ \tilde{\lambda } i}$ is a basis function in the
spin-isospin space with given $S$ and $T$ and belonging to
$\tilde{\lambda }$, the conjugate representation of $\lambda$.
Taking advantage of group theory, one has obtained the allowed
$\lambda $, which depends on $S$ and $T$\cite{Itzy}. The result in
the case of $T=1$ is listed in Table 1. We shall then figure out
which components are favorite to forming bound states.

\begin{table}[h]
\begin{center}
\caption{The allowed representation $\lambda $ of the states with
isospin $T=1$} \vspace*{2mm}
\begin{tabular}{|c|c|c|} \hline
$T$ & $S$ & $\lambda $  \\ \hline
 1 &  0  & $\{2\,1^4\}$, $\{2\, 2\, 2\}$, $\{3\, 2\, 1\}$, $\{3\,1^3\}$,
$\{4\, 2\}$ \\ \hline
 1 & 1 & $\{1^6\}$, $\{2\,1^4\}$, $2\{2\,2\,1\,1\}$, $\{3\,1^3\}$,
$2\{3\,2\,1\}$, $\{3\,3\}$, $\{4\,1\,1\}$ \\ \hline
 1 & 2 & $\{2\,1^4\}$, $\{2\,2\,1\,1\}$, $\{2\,2\,2\}$, $\{3\,1^3\}$,
$\{3\,2\,1\}$ \\ \hline
1  & 3 & $\{2\,2\,1\,1\}$ \\
\hline
\end{tabular}
\end{center}
\end{table}

For the function $F_{LSM}^{\lambda i}$ of the spatial coordinates,
defining a body-frame, we have
\begin{equation}
F_{LSM}^{\lambda i}(123456)=\sum_Q D_{QM}^L(-\gamma ,-\beta
,-\alpha )F_{LSQ}^{\lambda i}(1^{\prime }2^{\prime }3^{\prime
}4^{\prime }5^{\prime }6^{\prime }) \, ,
\end{equation}
where $\alpha$, $\beta$, $\gamma$ are the Euler angles to specify
the collective rotation, $D_{QM}^L$ is the well known Wigner
function, $Q$ is the projection of $L$ along the
$k^{\prime}$-axis, the $(123456)$ and $(1'2'3'4'5'6')$ specify
that the coordinates are relative to the laboratory frame and the
body-frame, respectively.

Since the $F_{LSQ}^{\lambda i}$ spans a set of basis of the
representation of the rotation group, space inversion group, and
permutation group, the invariance of the OCTA to the operations
$\hat{O}_1$ to $\hat{O}_4$ leads to four sets of equations. For
example, from
\begin{equation}
\hat{O}_1 F_{LSQ}^{\lambda i}({\cal{A}})=F_{LSQ}^{\lambda
i}(\hat{O}_1{\cal{A}})=F_{LSQ}^{\lambda i}({\cal{A}}) \, ,
\end{equation}
where $F_{LSQ}^{\lambda i}({\cal{A}})$ denotes that the
coordinates in $F_{LSQ}^{\lambda i}$ are given at an OCTA, for all
$Q$ with $|Q|\leq L$ we have
\begin{equation}
\sum_{i^{\prime }}[g_{ii^{\prime }}^\lambda (P(1432))e^{-i\frac
\pi 2Q}-\delta _{ii^{\prime }}]F_{LSQ}^{\lambda i^{\prime
}}({\cal{A}})=0 \, ,
\end{equation}
where $g_{ii^{\prime }}^\lambda $ is the matrix element belonging
to the representation $\lambda $, which can be fixed with the
group theory method (see for example Ref.\cite{group}). Similarly,
for $\hat{O}_2$, $\hat{O}_3$ and $\hat{O}_4,$ we have
\begin{equation}
\sum_{Q^{\prime }i^{\prime }}[g_{ii^{\prime }}^\lambda
[P(253)P(146)]\sum_{Q^{\prime \prime }}D_{QQ''}^L(0,\theta
,0)e^{-i\frac{2\pi }3Q''}D_{Q^{\prime }Q''}^L(0,\theta ,0)-\delta
_{ii^{\prime }}\delta _{QQ^{\prime }}]F_{LSQ^{\prime }}^{\lambda
i^{\prime }}({\cal{A}})=0 \, ,
\end{equation}

\begin{equation}
\sum_{Q^{\prime }i^{\prime }}[(-1)^Lg_{ii^{\prime }}^\lambda
(P_{14}P_{23}P_{56})\delta _{\bar{Q}Q^{\prime }}-\delta
_{ii^{\prime }}\delta _{QQ^{\prime }}]F_{LSQ^{\prime }}^{\lambda
i^{\prime }}({\cal{A}})=0 \, ,
\end{equation}
with $\bar{Q}=-Q$,
\begin{equation}
\sum_{i^{\prime }}[g_{ii^{\prime }}^\lambda
(P_{13}P_{24}P_{56})(-1)^{L} -\delta _{ii^{\prime
}}]F_{LSQ}^{\lambda i^{\prime }}({\cal{A}})=0 \, .
\end{equation}

Eqs.~(10) to (13) are the equations that the $F_{LSQ}^{\lambda
i}({\cal{A}})$ have to fulfill. In some cases there is one or more
than one nonzero solution(s) to all these equations. In some other
cases, however, there are no nonzero solutions. In the latter
case, the $\Psi _{LS}$ has to be zero at the OCTA configurations
disregarding their size and orientation. Accordingly, an INS
emerges and the OCTA is not accessible. Evidently, the solution of
above equations depends on and only on L, $\pi $ and $\lambda$.
Therefore the existence of the INS does not depend on the dynamics
(e.g., not on the interaction, mass, etc.) at all.

Solving the sets of linear equations, we obtain the accessibility
of the symmetry configurations of the OCTA with $L=0$. The result
is listed in the second row of Table 2, where the numbers in the
blocks are the ones of the independent nonzero solutions.

\begin{table}[h]
\begin{center}
\caption{\small The accessibility of the OCTA and the C-PENTA to
the $L^{\pi} =0^{+}$ wavefunctions with different spatial
permutation symmetries $\lambda $.} \vspace*{2mm}
\setlength{\tabcolsep}{1mm}
\begin{tabular}{|c|c|c|c|c|c|c|c|c|c|c|c|} \hline
 $\lambda $ & \{6\} & \{5\,1\} & \{4\,2\} & \{4\,1\,1\} &
 \{3\,3\} & \{3\,2\,1\} & $\{3\,1^3\}$ & \{2\,2\,2\} &
 \{2\,2\,1\,1\} &  $\{2\, 1^4\}$ & $\{1^6\}$          \\ \hline
 OCTA  & 1 & 0 & 1 & 0 & 0 & 0 & 0 & 1 & 0 & 0 & 0    \\ \hline
 C-PENTA & 1 & 1 & 1 & 0 & 1 & 2 & 0 & 1 & 1 & 1 & 1  \\  \hline
\end{tabular}

\end{center}
\end{table}

The INS existing at the OCTA may extend beyond the OCTA. For
example, when the shape in Fig.~1 is prolonged along $k'$ (called
a prolonged-octahedron and denoted as ${\cal{B}}$ ), it is
invariant to $\hat{O}_1,\, \hat{O}_3,$ and $\hat{O}_4$, but not to
$\hat{O}_2$. Hence, the $F_{LSQ^{\prime }}^{\lambda i^{\prime }}
({\cal{B}})$ should fulfill Eqs.(10), (12) and (13). When nonzero
common solutions of Eqs.~(10) to (13) do not exist, the INS
extends from the OCTA to the prolonged-octahedrons. In fact, an
OCTA has many ways to deform, for instance, instead of a square,
the particles 1, 2, 3 and 4 form a rectangle or form a diamond,
and so on. Then, the INS at the OCTA has many possibilities to
extend. How it extends is determined by the ($L^{\pi} \lambda$) of
the wavefunction. Thus, in the coordinate space, the OCTA\ is a
source for the INS to emerge. For a wavefunction, if the OCTA is
accessible, all the shapes in the neighborhood of the OCTA are
also accessible. Therefore this wavefunction is inherent nodeless
in this domain.

Another shape with a strong geometric symmetry is the regular
pentagon pyramid (PENTA, see Fig.~2). In an extreme case, the
PENTA can be C-PENTA, which corresponds to that with $h=0$ in
Fig.~2. Let $k'$ be the 5-fold axis, the C-PENTA is invariant to
(i) a rotation about $k'$ by $\frac{2\pi }5$ together with a
cyclic permutation of the five particles of the pentagon , (ii) a
rotation about $k'$ by $\pi $ together with a space inversion,
(iii) a rotation about $i'$ by $\pi $ together with $P_{14}P_{23}$
(here $i'$ is the axis vertical to $k'$ and connecting O and
particle 5). These invariances lead to constraints embodied in a
set of homogeneous equations, and therefore the accessibility of
the C-PENTA can be identified as given in the third row of Table
2.

In addition to the OCTA, the C-PENTA is another source where the
INS may emerge and extend to its neighborhood; e.g., extend to the
pentagon-pyramid as shown in Fig.~2 with h$\neq $0. There are also
other sources. For example, the one at triangular-prism and that
at the regular hexagons. However, among the 15 bonds, 12 can be
optimized at an OCTA, 10 at a pentagon-pyramid, 9 at a
triangular-prism and 6 at a hexagon. Therefore in the neighborhood
of the hexagon (and also other regular shapes) the total potential
energy is considerably higher. Since the wavefunctions of the
low-lying states are mainly distributed in the domain with a lower
potential energy, we shall concentrate only in the domains
surrounding the OCTA\ and the C-PENTA.

In most cases, the ground state of a nucleus obeys the condition
$T=T_3$. Thus, we can only consider $T=1$ instead of $T_3=1$, if
we constrict our discussion in the low-lying states.

Referring to Table 2, one can find that, when a $\Psi_{LS\lambda}$
has $(L^{\pi} \lambda ) =(0^+\{6\})$, $(0^+ \{4\,2\})$, or
$(0^+\{2\,2\,2\})$, it can access both the OCTA and the C-PENTA.
These and only these wavefunctions are inherent-nodeless in the
two most important domains, and they should be the dominant
components for the low-lying states. All the other $L=0$
components must contain at least one INS. Table 1 shows that the
($0^+\{6\}$) component can not be contained in any $T=1$ state.
Then, we obtain that the ($0^+ \{4\,2\}$) component is accessible
to $[T, S]=[1, 0]$ state, and the ($0^+ \{2\,2\,2\}$) component is
allowed to $[T, S]=[1, 0]$ and $[1, 2]$ states.

Since the state with $[T, S]=[1, 0]$ is accessible to both the
$\{4\, 2\}$ and $\{2\, 2\, 2\}$ components,  two $J^{\pi} =0^{+}$
partner-states would be generated. Each of them is mainly a
specific mixture of the $\{4\, 2\}$ and $\{2\, 2\, 2\}$
components. When $[T, S]=[1, 2]$, there is only one accessibility.
We have then only one $J^{\pi} =2^{+}$ state. Therefore, we
predict that the low-lying $T=1$ positive parity spectrum of a
6-nucleon system involves totally three states. Two of them have
total angular momentum $J^{\pi} = 0^{+}$, and another one has
$J^{\pi} = 2^{+}$. All the quantum numbers of these states are
listed in Table 3.

According to experiment data, besides positive parity states, some
low-lying resonances with negative parity have also been observed
in light nuclei. Therefore, in addition to the $L=0$ case
discussed above, we have to study the case of $L=1$.

By evaluating the determinants of the sets of homogeneous linear
equations, the inherent nodeless components of a nucleus with 6
nucleons and  $L^{\pi}=1^{-}$ are identified as the ones holding
orbital symmetry
\begin{equation}
\lambda  \in \{ \{5\, 1\}, \{4\, 1\, 1\}, \{3\, 3\}, \{3\, 2\,
1\}, \{2\, 2\, 1\, 1\} \}\, .
\end{equation}
Since these states have angular momentum $L=1$, the total angular
momentum $J$, which is formed by the coupling of $S$ and $L$, have
always three choices if $S\ne 0$. From Table 1 one knows that the
$\{5 \, 1\}$ component is not allowed to the $T=1$ states. Then,
there exist three groups of P-wave states each with spin $S=0, 1,
2$, respectively. Their quantum numbers and orbital symmetries are
listed in Table 3, too.

\begin{table}[h]
\begin{center}
\caption{Prediction of the quantum numbers of low-lying states of
the $T=1$ six-nucleon systems and the energies of the states of
$^6$He in experiments(the data marked with a, b, c are taken from
Refs.\cite{Ajzen}, \cite{Janecke}, \cite{Naka}, respectively).}
\vspace*{2mm}
\begin{tabular}{|c|c|c|c|c|c|c|c|}

\hline $T$ & $S$ & $J$ & $\pi $ & $L$ & $\lambda $ & E$_x$/MeV & $\Gamma$/MeV  \\
\hline
1 & 0 & 0 & $+$ & 0 & \{4\,2\}, \{2\,2\,2\} & $0^{a}$ &  \\
\hline 1 & 2 & 2 & $+$ & 0 & \{2\,2\,2\}  & $1.797^{a}$  &  \\
\hline 1 & 0 & 0 & $+$ & 0 & \{4\,2\}, \{2\,2\,2\}& $5.6\pm
0.6^{b}$ &  $10.9 \pm 1.9^{b}$ \\
\hline 1 & 2 & 2 & + & 0 &  & $5.6 \pm 0.6^{b}$  & $10.9 \pm 1.9^{b}$ \\
\hline 1 & 0 & 1 & $-$ & 1 & \{3\,2\,1\} & $4\pm 1^{c}$  & $4\pm 1^{c}$ \\
\hline 1 & 1 & 0, 1, 2 & $-$ & 1 & \{4\,1\,1\}, \{3\,3\},
\{3\,2\,1\}, \{2\,2\,1\,1\} & $14.6 \pm 0.7^{b}$ & $7.4 \pm 1.0^{b}$ \\
\hline 1 & 2 & 1, 2, 3, & $-$ & 1 & \{3\,2\,1\}, \{2\,2\,1\,1\} &
$14.6 \pm  0.7^{b}$ & $7.4 \pm 1.0^{b}$ \\
\hline

\end{tabular}
\end{center}
\end{table}

It has been well known that the states having two or more
components will split due to the coupling among them. For the
positive parity states, one can obtain that, owing to the
interference between the $\{4\, 2\}$ and $\{2\, 2\, 2\}$
components, there would be a large energy gap between the two
$J^{\pi} = 0^{+}$ partner-states, so that the lower one becomes
the ground state, while another has an energy higher than the
first excited state $2^+$. Recalling the energy spectrum of $^6$Li
whose gap between the ground state and its partner is $5.65$~MeV
and considering the charge independent characteristic of nuclear
force and the similarity of the nuclei $^{6}$He and $^{6}$Li, we
expect that the $0^+_2$ state of $^6$He would have an energy
$E_x=5.6$~MeV. Meanwhile, the expected $J^\pi= 2^+\; ([T,S]= [1,
2])$ state is the first excited state at $E_x=1.797$~MeV. For the
negative parity states (P-wave resonance), the $J^{\pi} = 1^{-}$
state with $[T,S] =[1, 0]$ and orbital symmetry $\{3\, 2\, 1\}$
would be the lowest one. The resonance with $J^{\pi} = 2^{-}$,
$1^{-}$, $0^{-}$ and $[T,S] = [1, 1]$ and those with $J^{\pi} =
3^{-}$, $2^{-}$, $1^{-}$ and $[T,S]= [1,2]$ are the second and the
third set of resonant states, respectively. In experiments, the
lowest $1^{-}$ resonance once was assigned as a component of the
resonance at $E_x = 5.6$~MeV with a width $\Gamma = 10.9$~MeV in
the $^{6}$Li($^{7}$Li, $^{7}$Be)$^{6}$He reaction\cite{Janecke}.
The subsequent similar experiment provided a clue that it would be
the one at $E_x=4 \pm 1$~MeV with $\Gamma = 4 \pm
1$~MeV\cite{Naka}. While the recent $^{6}$Li(t, $^{3}$He)$^{6}$He
experiment\cite{Nakamura} indicated that the broad resonance at
$E_x \approx 5.6$~MeV involves at least three Gaussians with
energy $4.4\pm 0.1$~MeV, $7.7\pm 0.2$~MeV, $9.9\pm 0.4$~MeV,
respectively. And each of these three peaks contains the $1^{-}_1$
and other components. In some detail, the other constituents of
this broad resonance must involve a $2^{+}$
state\cite{Janecke,Nakamura}. Even though the present approach can
not give this state naturally since it is not a nodeless
accessible one(then we leave the configuration $\lambda$ in Table
3 blank), we can assign it as the one dominated by the
PENTA-accessible but OCTA-inaccessible S-wave components $\{3\,
2\, 1\}$ and $\{2\, 2 \, 1 \, 1\}$, even $\{2\, 1^{4} \}$.  In a
short word,  the resonance centred at $5.6$~MeV contains at least
the $0^{+}_2$, $1^{-}_1$ and $2^{+}_2$ states. As to the $1^{-}_2$
and $2^{-}_1$ states, according to the $^{6}$Li($^{7}$Li,
$^{7}$Be)$^{6}$He experiment\cite{Janecke}, they correspond to the
resonance at $E_{x}= 14.6 \pm 0.7$~MeV with $\Gamma = 7.4\pm
1.0$~MeV. If the possible splitting of this resonance into two or
three components in the range $E_x = 13 \sim 18$~MeV proposed in
Ref.\cite{Janecke} is confirmed, the $J^{\pi} = 0^{-}$ resonance
may also be included. In addition, the $3^{-}$ and other $2^{-}$,
$1^{-}$ resonance may be the constituents of that at $E_x \approx
23.3$~MeV reported in Ref.\cite{Janecke}. Summarizing these
experimental correspondences, we list the data in Table~3 and
illustrate the energy spectrum in Fig.~3. From Table~3 and Fig.~3,
one can recognize that the energy spectrum of $^{6}$He obtained in
the present analysis agrees very well with experimental data (for
a recent compilation, see Ref.\cite{Til02}).

Since $^6$Be and $^6$He are mirror nuclei, each of which has $A =
6$ and $|T_3|=1$, we can extend the discussion above to induce the
quantum numbers of the low-lying states of $^6$Be and obtain the
same result as for $^{6}$He. Unfortunately, only two low-lying
states have been observed in experiments\cite{Til02}. For
comparison we illustrate these states in Fig.~3, too.

In summary, using the inherent nodal surface structure analysis,
we have determined the quantum numbers of the low-lying states of
6-nucleon systems with isospin $T=1$. The orbital symmetries $\{4
\, 2\}$ and $\{2\, 2\, 2\}$ are found to be the important
components for the S-wave states and the $\{3\, 2\, 1\}$, $\{2\,
2\, 1\, 1\}$ symmetries for the P-wave states. The energy spectra
of $^{6}$He and $^{6}$Be obtained in the present analysis agree
very well with the experimental data, except for the $2^{+}_{2}$
state. In fact, although the $2^+_{2}$ state does not appear as a
low-lying one in our result, we haven't excluded its possibility
to be resonance at higher energy. On the other hand, a shell model
calculation\cite{DRE97} once predicted that the second $0^+$ state
of $^{6}$He is at $12$~MeV. However, in our analysis, this state
must be much lower, at least as low as $1^{-}_1$ and $2^{+}_2$
states, which is quite consistent with the available experimental
data at present. The present result provides then further evidence
for that the inherent nodal surface analysis is a quite powerful
approach for few-body problems.

It is evident that, although our analysis is simply based on the
INS analysis, the obtained energy level structure agrees quite
well with the experimental data.  It indicates that the INS
embodies the basic and intrinsic properties of the system (e.g.,
symmetry, configuration's structure) and with which one can
discuss both positive and negative parity states simultaneously.
Thus, it can help us to make reasonable choice between dynamical
models. Meanwhile, the inherent nodeless wave functions are the
most important building blocks of the low-lying states. The
identification of these favorite components plays then a key role
in understanding the low-lying energy spectrum. However, it is
remarkable that the INS analysis can not give accurate numerical
results directly. To get numerical results one must implement
dynamical calculations. Then combining dynamical calculation and
the INS analysis is the efficient way to investigate few-body
problems.
\bigskip

This work is supported by the National Natural Science Foundation
of China with Grant No. 10075002, 10135030, 19875001, 90103028,
the Major State Basic Research Developing Programme under Grant
No. G2000077400 and the Foundation for University Key Teacher by
the Ministry of Education, China.

\newpage



\begin{center}
\begin{figure}
\includegraphics[scale=0.65,angle=0]{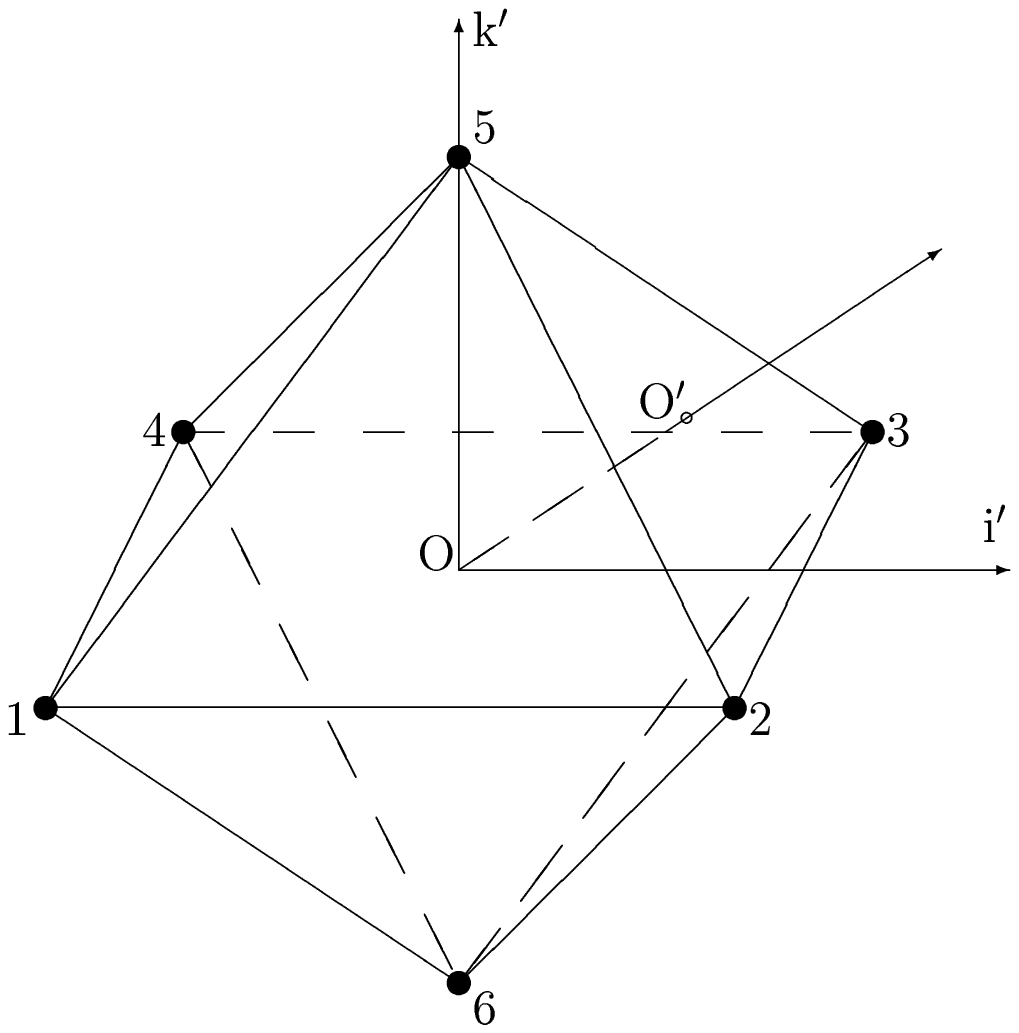}
\caption{The sketch of regular octahedron }
\end{figure}
\end{center}

\begin{center}
\begin{figure}
\includegraphics[scale=0.65,angle=0]{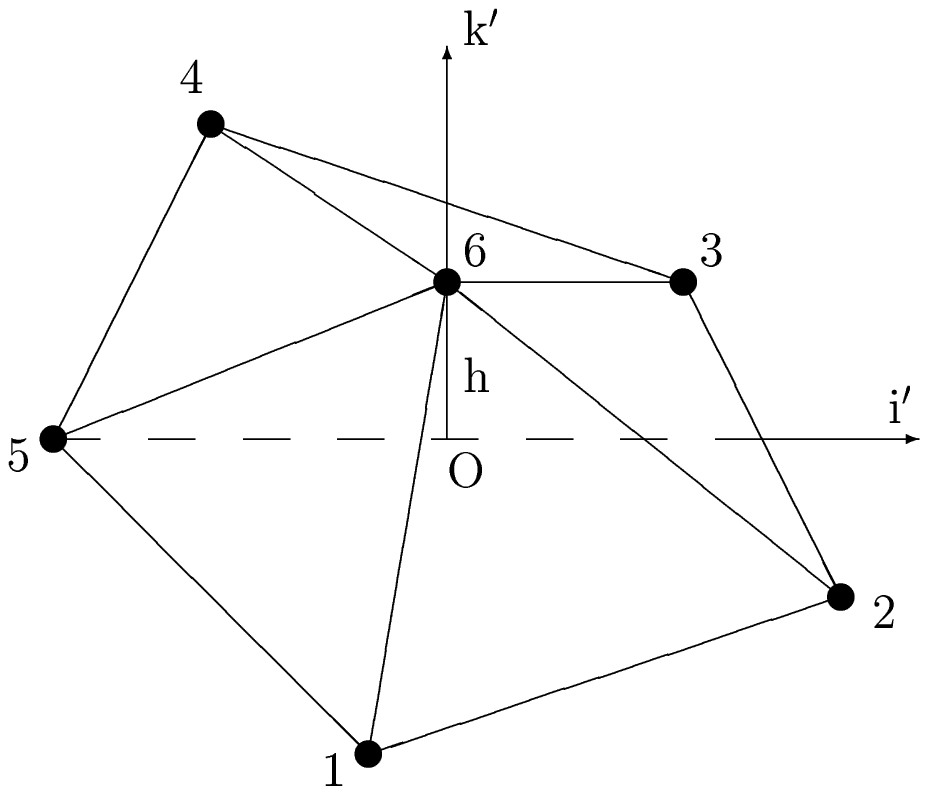}
\caption{The sketch of regular pentagon pyramid }
\end{figure}
\end{center}

\begin{center}
\begin{figure}
\includegraphics[scale=0.65,angle=0]{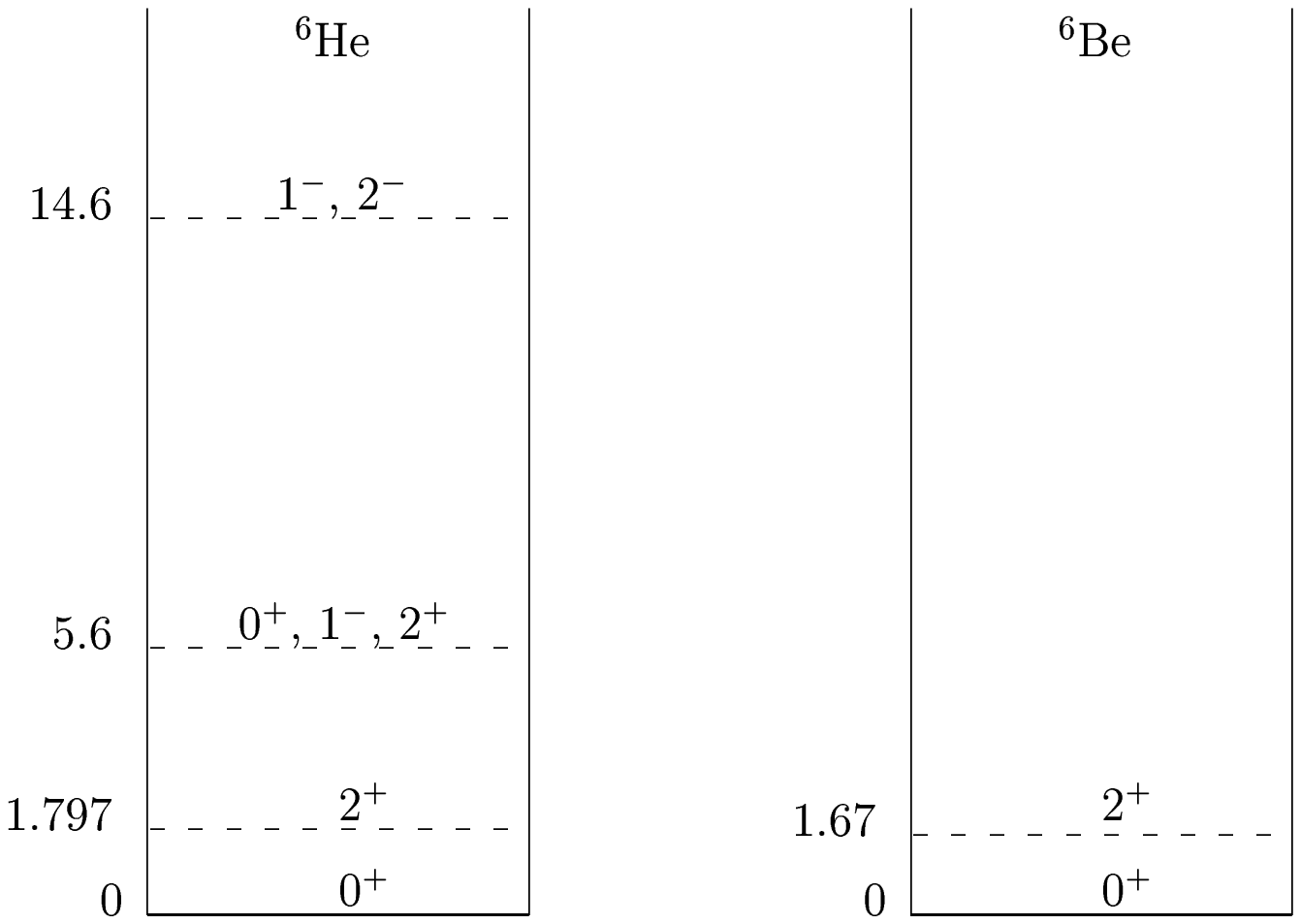}
\caption{The energy level scheme of $^6$He and $^6$Be }
\end{figure}
\end{center}

\end{document}